# MeV Argon ion beam generation with narrow energy spread

Jiancai Xu[1], Tongjun Xu[1], Baifei Shen[1,2,*], Hui Zhang[1], Shun Li[1], Yong Yu[1], Jinfeng Li[1], Xiaoming Lu[1], Cheng Wang[1], Xinliang Wang[1], Xiaoyan Liang[1,2], Yuxin Leng[1,2], Ruxin Li[1], and Zhizhan Xu[1]

[1]*State Key Laboratory of High Field Laser Physics, Shanghai Institute of Optics and Fine Mechanics, Chinese Academy of Sciences, P. O. Box 800-211, Shanghai 201800, China*

[2]*Collaborative Innovation Center of IFSA (CICIFSA), Shanghai Jiao Tong University, Shanghai 200240, China*

**Laser driven particle acceleration has shown remarkable progresses in generating multi-GeV electron bunches and 10's of MeV ion beams based on high-power laser facilities. Intense laser pulse offers the acceleration field of TV/m ($10^{12}$V/m), several orders of magnitude larger than that in conventional accelerators, enabling compact devices. Here we report that a highly-collimated argon ion beam with narrow energy spread is produced by irradiating a 45-fs fully-relativistic laser pulse onto an argon cluster target. The highly-charged ($Ar^{16+}$) heavy ion beam has a minimum absolute energy spread of 0.19 MeV/nucleon at the energy peak of 0.39 MeV/nucleon., we identify a novel scheme from particle-in-cell simulations that greatly reduces the beam energy spread. The laser-driven intense plasma wakefield has a strong modulation on the ion beam in a way that the low energy part is cut off. The pre-accelerated argon ion beam from Coulomb explosion thus becomes more mono-energetic and collimated.**

Conventional ion accelerators provide powerful ion beams for experiments in nuclear physics, atomic physics, and so on. The injected ion sources into accelerator require intense enough particles with high charge state and narrow energy spread[1]. Laser driven ion beam makes a strong candidate for the ion injection. They are of short pulse duration, high charge state, low energy and high current[2]. With dramatic development of laser technology, the laser intensity achieves the relativistic scale with pulse duration of few femtoseconds. The intense laser field irradiates the gaseous or solid target and transfers energy to electrons and ions. The ion beam energy reaches 10's of MeV [3] and the energy spread gets reduced to a few precent[4-10]. Despite these progresses, acceleration of high-Z ions with high charge state has been rare. Moreover, in various laser accelerating schemes, when high-Z solid foil is employed, the proton in the target surface gets accelerated preferentially. Thus the final generated ion beam has multispecies.

Gaseous target has been proposed to avoid the impurity of high-Z ion sources. Gas purity can easily reach as high as 99.99%. To further increase the laser energy absorption, gas-cluster target has been used. Laser-cluster interaction leads to high ionization, acceleration of hot electrons and highly-charged ions[11,12], and X-rays emission[13]. Ion beams with energy up to 1 MeV generated with cluster targets have been experimentally observed at the laser intensity of $10^{14}$-$10^{16}$ W/cm$^2$ [14,15]. The ions energy distribution usually fits the Coulomb explosion model[16,17], and the ion energy spectrum turns out to be very broad. However, certain applications such as cancer therapy[18] and injector of conventional ion accelerator[2] require the ion beam with low energy spread.

Here we have experimentally demonstrated the generation of high-quality argon ion beam from the interaction between a 45-fs fully-relativistic laser pulse and an argon cluster target. The generated argon ion beam carries a charge state as high as $Ar^{16+}$. It has a minimum Full-Width-Half-Maximum (FWHM) energy spread of 7.5 MeV, corresponding to 0.19 MeV/nucleon at the energy peak of 0.39 MeV/nucleon. In addition, the argon ion beam is highly collimated with a small divergence angle of 2.4º (FWHM).

*Author to whom correspondence should be addressed. Electronic mail: bfshen@mail.shcnc.ac.cn.

The experiments have been carried out by using the femtosecond petawatt laser system at Shanghai Institute of Optics and Fine Mechanics (SIOM), Chinese Academy of Sciences (CAS)[19]. Figure 1a describes the experimental setup (see Methods). A 45 fs fully-relativistic laser pulse irradiates an argon cluster target, and the high-energy electrons and argon ions are generated. Figure 1b is the obtained raw data for argon ion beam recorded by the image plate. The observed high-energy argon ions correspond to $Ar^{16+}$. It is centered between 20-30MeV with negligible low-energy ions, indicating efficient ion acceleration with low energy spread. After vertical integration and deconvolution, the corresponding argon ion spectrum is plotted in Fig. 2a at the case of $n = 0.25\,n_c$, where $n_c$ is the critical density for the laser central wavelength of 800nm. It has a peak value at 21.8 MeV, corresponding to 0.55 MeV/nucleon with a FWHM energy spread of 0.53 MeV/nucleon. The spectrum extends up to 80MeV. The vertical distribution of argon ion signal in Fig.1b describes the beam divergence, which indicates a highly collimated argon ion beam with a small divergence angle of 2.4º (FWHM). This is smaller by a factor of 4 than that of the ion beam generated by laser-solid interaction [20].

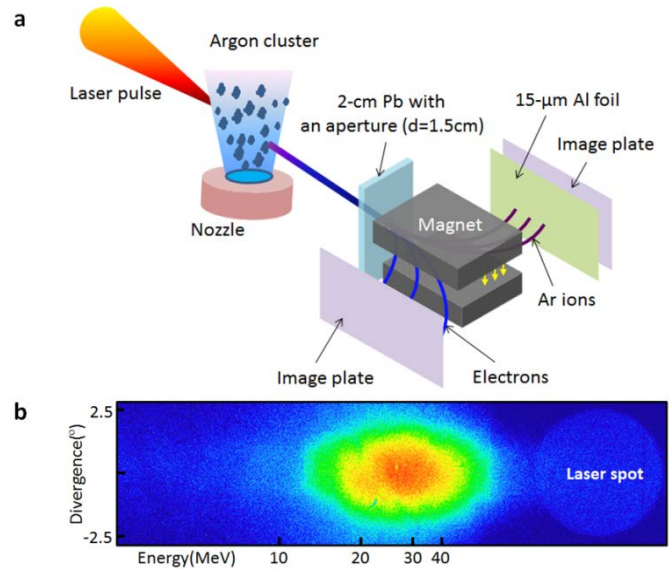

**Figure 1 a,** Experimental arrangement: An intense femtosecond laser pulse is focused into an argon cluster jet. The electrons and ions are accelerated to high energy, and propagate into a magnetic spectrometer. The entrance aperture is a 2cm-thick lead with a hole of 1.5 cm diameter, which limits the ion beam within the full divergence angle of 5º. **b,** Raw image of argon ion beam recorded in the image plate, where the backing pressure of the argon gas is 32bar. The vertical axis represents the beam divergence. The 'laser spot' region is the X-ray radiation from the plasma, and it represents the laser propagation axis.

High-quality argon ion beams with small energy spread have been observed for more than 20 shots in different argon gas conditions. An example of six spectra with the similar laser parameter but different background electron densities of $0.18 - 0.36 n_c$ is presented in Fig. 2a. The backing argon gas pressure varies from 23 to 47 bar as the laser pulse is focused in the argon cluster 500 μm away from the nozzle exit. All these experimental spectra show well-defined energy peaks with narrow energy spreads. The energy peak varies between 20 and 25 MeV except for the case of $n = 0.36n_c$, where the peak energy of the ion beam shifts to 38 MeV (0.95 MeV/nucleon). We attribute the higher peak energy at $n = 0.36n_c$ to the fact that larger-sized clusters are formed with the higher background density. Further experiments are on the way to confirm this. A surprisingly low energy-spread spectrum was observed at the backing gas pressure of 10.5bar, as shown in Fig. 2b. It has a greatly reduced energy spread of 7.5 MeV (FWHM), which is 0.19 MeV/nucleon at the energy peak of 0.39 MeV/nucleon, corresponding to the relative energy spread of 49%. The laser pulse focused in the argon cluster at a distance of 1.3mm from the nozzle exit. In the similar experimental condition, we observed the high-qualtiy argon beam with narrow energy spread in at least 4 shots, and the average energy spread for these shots is 16.9 MeV, corresponding to 0.42 MeV/nucleon. All these results show that the intense laser-cluster interaction has an ability to accelerate high-quality argon ion beam with low energy spread.

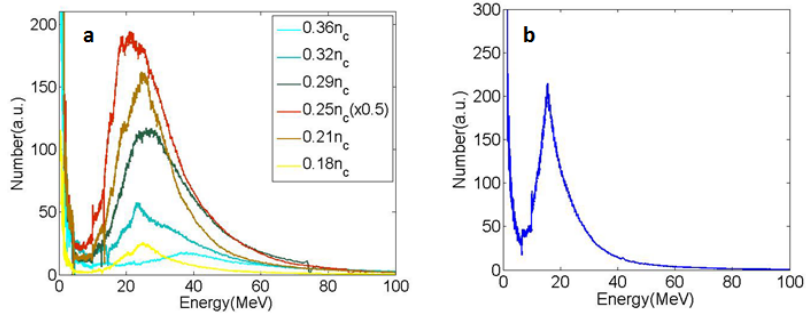

**Figure 2 a** Observed ion beam spectra in experiments versus the background electron density of 0.18—0.36$n_c$. The lineout of 0.25$n_c$ case is reduced by 2X to fit on the same scale and the used image plate for these shots is BAS-TR. The argon atom density was measured independently through optical interferometry method, and the electron density is calculated by assuming that the argon atom is ionized to $Ar^{16+}$. **b**, Monoenergetic argon ion beam observed with backing pressure of 10.5 bar. Here a 15 μm-thick aluminum foil covers the entrance of magnet instead of image plate, and the BAS-SR image plate is used for argon ion detection.

It has been demonstrated that laser-cluster interaction experiments at low laser intensity of $10^{14}$-$10^{16}$ W/cm$^2$ produced the high-charge ion beam due to Coulomb explosion[11,14]. When the laser intensity increases to the relativistic scale, the charge state of argon ion from the cluster target could increase up to $Ar^{16+}$ and the argon ion energy exceeds MeV[21]. However, the nature of Coulomb explosion always leads to a very broad ion spectrum[16]. The surprising narrow energy spread observed in our experiments indicates that the argon ion beam must have been somehow modulated through a unique and yet unexplored mechanism, which could greatly improve the beam quality. The novel mechanism, as we will reveal in the following, is the laser-driven wakefield modulation.

We identify this novel scheme based on two-dimensional particle-in-cell (PIC) simulations with the code VORPAL[22] (simulation details in Methods). High-quality argon ions with narrow energy spread were also observed in the simulation. This novel scheme is distinguished into two distinct stages, i.e. Coulomb explosion and wakefield modulation, as shown in Fig. 3. In the first stage, the intense laser field displaces the electrons of the argon cluster and induces Coulomb explosion at $t = t_1$ (Fig. 3a). The ions get energy gain during the ion expansion and thus have a flat continuous spectrum, as presented in Fig. 3b. Simultaneously, the intense laser pulse displaces the background electrons through the ponderomotive force, causes the charge separation, and drives a strong co-moving plasma wave—the laser wakefield[23]. Figure 3c displays the longitudinal periodic electric wakefield behind the laser pulse. Since the accelerated ions have much lower velocities than the group velocity of laser pulse in the underdense plasma, they all fall into the wakefield. At a later time $t = t_2$, different parts of the ion beam experience the acceleration and deceleration phase of the laser-driven wakefield. We come to the second stage—wakefield modulation. When the ion beam moves out of the wakefield at $t = t_3$, the forward-moving ions separate into two parts, high-energy part and low-energy part, which contribute to the high-energy peak and low-energy background in the ion spectrum shown in Fig. 3d, respectively. In this way, the laser wakefield significantly reduces the energy spread of the high-energy beam head by cutting off the low energy tail.

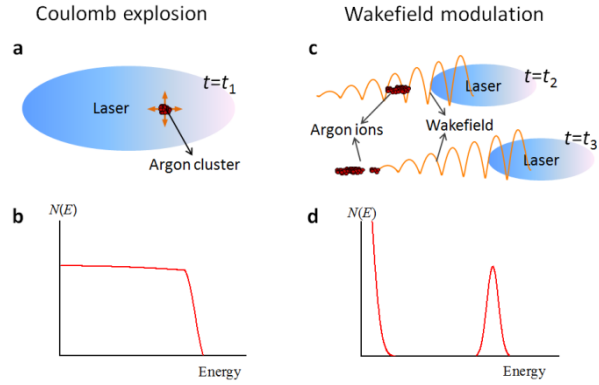

**Figure 3** Sketch of narrow-energy-spread ion beam generation in the interaction between a fully-relativistic laser pulse and an argon cluster target. There are two different stages: Coulomb explosion (**a, b**) and wakefield modulation(**c, d**). The laser pulse propagates from left to right in the argon plasma.

This new scheme was presented based on one argon cluster interacted with the laser pulse as well as the laser-driven wakefield. To show the modulation mechanism of the argon ion in more details, we plotted in Fig. 4 the ions momenta $P_Z$ and the electric field $E_Z$ on the propagation axis $z$, where the laser pulse propagates from left to right in the plasma. At $t = 30T_0$, the laser peak locates at $z = 12$μm while the argon cluster is set at $z = 15$μm. The laser front irradiates the cluster, expels all electrons and creates a strong charge separation field. The argon ions start to explode. As shown in Fig. 4a, the ions momentum $P_Z$ has a linear dependence on the longitudinal position $z$, which is an obvious signature of Coulomb explosion[21]. At $t = 40T_0$, the laser pulse peak intensity passes over the cluster, and the corresponding ion spectrum, as plotted in Fig. 4e, has a continuous flat spectrum until the cutoff energy ($E_{\text{cutoff}} = 20$ MeV). This is the typical spectrum of accelerated ions due to the Coulomb explosion. During the ion expansion, the ion spatial distribution has the 'donut' structure (see supplementary Fig. 1), which is different from the Coulomb explosion induced by low-intensity laser pulse [24].

At the same time of displacing the cluster electrons, the laser pulse also expels the background electrons and drives a plasma wakefield in background plasma. The electric field $E_Z$ co-propagates with the laser pulse. The Coulomb explosion provides the pre-acceleration of argon ions, but the ions' velocities are still much smaller than the group velocity of the laser pulse in the underdense plasma. Therefore each ion would experience multi-periodic wakefield $E_Z$, where in each period the electric field $E_Z$ is positive in the first half and negative in the second half. At

$t = 60T_0$, the electric field $E_z$ has two periods in the window size, which shows only a fraction of the simulation box. The whole cluster ions occupy more than half period of the electric field $E_z$. Created by the laser leading Coulomb explosion, the front half part of ions with high initial momenta $P_z$ experience the acceleration phase of the electric field $E_z$ ($E_z > 0$) and get further acceleration. But the relatively low energy ions are decelerated by the negative wakefield. This modulation process can be clearly seen in Fig.4b.

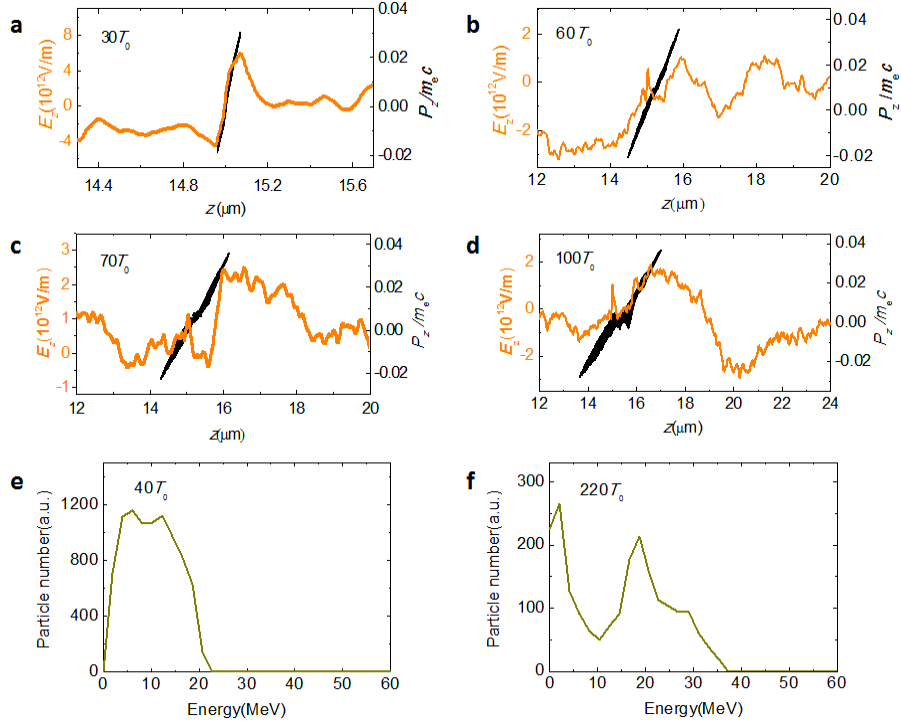

Figure 4 Argon ions spectrum modulation by the laser-driven longitudinal field: the argon ion momenta $P_z$ and the electric field $E_z$ in the longitudinal direction (**a-d**) as well as the ion spectra (**e, f**) at different time steps. The ions momenta $P_z$ are normalized to $m_e c$, where $m_e$ is the electron mass, and $c$ is the light speed in vacuum. The laser pulse propagates from left to right in the plasma and the window only shows a fraction of the simulation box.

Similar modulation process is also seen during the subsequent wakefield periods in Fig. 4c and d. Since the field is strongest in the first few plasma periods and decays rapids behinds, it is the first few wakefield structures that play the most important role in the modulation. The consequence of the wakefield modulation is that the high-energy beam head propagates faster than the rest of the beam and finally they get separated in space. The wakefield here poses a barrier on the ion energy that filters the low energy ions and narrows the energy spread of the high energy ones, as shown in Fig. 4e and f. The broad ion spectrum at $t = 40T_0$ has been narrowed down to an absolute energy spread of 8.3 MeV at the peak of 20 MeV. The post-acceleration process from the electric field $E_z$ increases the ions cut-off energy from 21 to 37 MeV. These results explain our experimental results very well.

PIC simulation confirmed that the fully-relativistic laser pulse drives strong periodic electric field, which efficiently modulates the continuous ion spectrum from Coulomb explosion in a way that the low energy part is cut off and thus brings the narrow-energy-spread ion beam. Here the laser wakefield has to be intense enough to modulate the ion spectrum. A comparative simulation was also performed without background plasma. It is found that the outgoing ion spectrum in the simulation has a plateau distribution with a 100% energy spread when the laser wakefield is absent (see Supplementary Fig. 2). This further proves the fact of the laser wakefield modulation.

We believe that the results presented here are not only the first observation of wakefield modulation on MeV laser-driven ion source but also provide an universal method to create highly-charged pure heavy ion beams with low energy spread. The employed cluster-gas target is not limited to argon but available for various ion species including Xe, Kr, C, O, N, and so on. We would like to point out that at current experimental condition, the ions experience all plasma periods and the ion spectrum gets strongly modulated and dramatically reduced by the laser-driven wakefield. At even higher laser intensities, the laser wakefield will offer such strong acceleration field to the pre-accelerated ions in the first plasma period that the ions will get trapped and further accelerated for a long distance[25]. And hence we move from the modulation regime to the acceleration regime.

In conclusion, a highly-charged argon ion beam ($Ar^{16+}$) with a small absolute energy spread of 0.19MeV/nucleon is generated experimentally in the interaction between a 45-fs fully-relativistic laser pulse and an argon cluster target. A novel scheme to obtain a high-quality ion beam with narrow energy spread has been presented and verified by the two-dimensional PIC simulation. The intense periodic electric field of laser wakefield has a strong modulation on the ion spectrum and thus significantly reduces the ion energy spread after the ions get energy gain from Coulomb explosion. The cluster-gas target can work on a high repetition rate. Since the 10Hz PW laser facility is in progress, the reproducible high-quality ion source with low energy spread will become more feasible for many applications in the near future, including medical therapy[18], fusion targets diagnostics[26], nuclear physics and injector of conventional ion accelerators.

**Methods**

**Laser and Argon ion beam characterization**:

The linearly-polarized laser pulse with the central wavelength of 800nm delivers the energy of 14 J on target within the pulse duration of 45 fs (FWHM) in single-shot working mode per half hour. An F/4 off-axis parabola mirror focuses the laser pulse into an argon cluster jet produced by the supersonic conical nozzle with an exit diameter of 650μm. The peak laser intensity achieves $7.2 \times 10^{19}$ W/cm$^2$ with a focus size of 12μm, corresponding to a normalized vector potential $a_0$=5.7. The pulsed nozzle works at room temperature (296K) with a pressure adjustment range from 10 to 50 bar. The argon cluster forms during the supersonic expansion of the argon gas with the purity of 99.999% in vacuum. For a certain nozzle used in experiments, the mean size of argon clusters varies at different backing pressures[27].

With the laser contrast of $10^{-8}$, the pre-pulse intensity is smaller than $10^{12}$W/cm$^2$. The pre-pulse has not destroyed the majority of the argon cluster before the main pulse arrives. The main pulse with high intensity irradiates the cold and overdense cluster, ionizes the argon atoms, and accelerates argon ions as well as electrons. High-energy electrons and ions pass through the entrance of magnet and then they are separated by a dipole permanent magnet spectrometer, which provides an effective magnetic field of 0.8 T over 15 cm with the gap of 4 cm. The entrance aperture is 2cm-thick lead with a hole with the diameter of 1.5 cm, which limits the ion beam divergence of 5º. The image plates detect electrons and ions simultaneously for each shot. In our experiments, two types of image plates (Fujifilm, BAS-SR and -TR) have been used. The image plate of BAS-SR has a 6-μm protective plastic layer in front of the phosphor layer, while the image plate of BAS-TR has no protective layer, where the argon ions irradiate the phosphor layer directly. A 15 μm-thick aluminum foil covers the image plate at the position of laser spot and ion part to block the laser pulse as well as plasma radiation, and background low-energy X-rays.

The maximum charge state of argon ion is $Ar^{16+}$ in our experiments, based on the optical-field ionization experiments [28]. It is much more difficult for further ionization as the ionization to $Ar^{17+}$ needs the energy of 4.12 keV[29], which cannot be realized in our experiments. The ions with higher charge states get larger energy gain in the same acceleration field, and thus the energy distribution of accelerated ions shifts to higher energy side when the charge state rises[21]. Therefore the high-energy argon ion recorded in image plate corresponds to $Ar^{16+}$.

**PIC simulation**: Two-dimensional PIC simulations were performed to investigate to the argon ion beam generation with low energy spread. One argon cluster is irradiated by a fully-relativistic laser pulse in the background argon plasma. The simulation box is 40 μm × 40 μm, divided into 8000× 5000 cells. The laser pulse has a temporal Gaussian profile with a duration of 40fs (FWHM), the normalized intensity of $a_0 = 8$, and the focus size of 12 μm in the transversal direction. The simulation does not include the ionization process. A pre-ionized argon cluster ($Ar^{16+}$) with a radius of 9 nm is set at $z = 15$ μm in the homogenous electron and argon ion plasma. The cluster density is 218 $n_c$. The background argon plasma ($Ar^{16+}$) has a homogenous distribution of $n = 0.3n_c$ in the range of 4<z<38 μm. There are 60000 macro particles per cell to describe the argon cluster while one macro particle per cell is set for the background argon plasma.

**Acknowledgements**

The authors are thankful to Dr. Liangliang Ji for helpful discussions and suggestions. This work was supported by Natural Science Foundation of China (Projects Nos.11125526, 11335013, 11505264, 61521093).

**Contributions**

J. Xu, T. Xu, B. Shen, S. Li, and Y. Yu designed and carried out the experiments. J. Xu and T. Xu did the data analysis. J. Xu and H. zhang performed the simulations. J. Xu and B. shen wrote the paper. All authors commented on the manuscript.

SUPPLEMENTARY INFORMATION

# MeV Argon ion beam generation with narrow energy spread

Jiancai Xu[1], Tongjun Xu[1], Baifei Shen[1,2,*], Hui Zhang[1], Shun Li[1], Yong Yu[1], Jinfeng Li[1], Xiaoming Lu[1], Cheng Wang[1], Xinliang Wang[1], Xiaoyan Liang[1,2], Yuxin Leng[1,2], Ruxin Li[1], and Zhizhan Xu[1]

[1]State Key Laboratory of High Field Laser Physics, Shanghai Institute of Optics and Fine Mechanics, Chinese Academy of Sciences, P. O. Box 800-211, Shanghai 201800, China
[2]Collaborative Innovation Center of IFSA (CICIFSA), Shanghai Jiao Tong University, Shanghai 200240, China

1. **Ion spatial distribution during the Coulomb explosion and wakefield modulation**

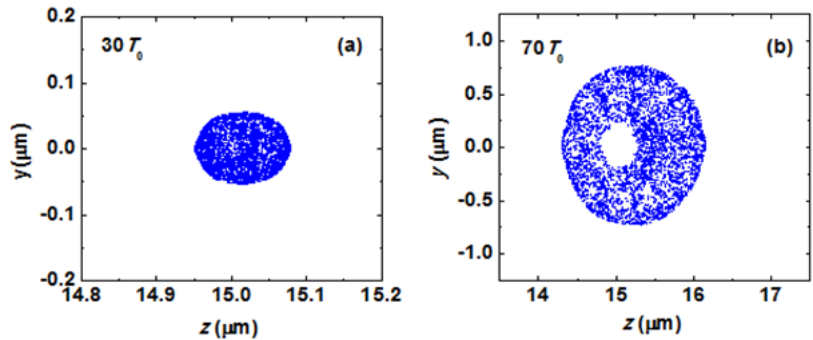

Supplementary Figure-1 Spatial distribution of argon ions at different time steps of $30T_0$ and $70T_0$.

Supplementary Fig. 1 describes the ion spatial distribution in the same simulation in Figure 4. At $t = 30T_0$, the laser intensity peak is located at $z = 12$μm while the argon cluster is set at $z = 15$μm, and the laser front irradiates the cluster, displaces sufficiently all the cluster electrons in the very short time, and creates the large positive space charge. The charge separation field is so strong that all ions are moving away from the cluster core, generating a hole in the middle. It can be clearly seen that the spatial distribution of ions has a typical 'donut' shape with a diameter of 700nm at $t = 70T_0$. Moreover, the ion expanding is not symmetric and the number of forward-moving ions is larger than that of back-moving ions as the wakefield accelerates the ions.

2. **Laser induced Coulomb explosion without background plasma**

In order to confirm the laser wakefield modulation on the ion spectrum, a comparative simulation without the background plasma was performed. All parameters of the laser pulse and the argon cluster were set as same as that in the simulation shown in Fig. 4. The intense laser field displaces all the cluster electrons and induces the strong Coulomb explosion. The ions have a 'donut' shape in the spatial distribution and they undergo symmetric Coulomb explosion. As described in Supplementary Fig. 2, the ion spectrum at $t = 40T_0$ is continuous with a small peak at 17MeV. After the ions expansion, the maximum energy of ions stops at 30MeV and the ion spectrum has a plateau distribution a 100% energy spread at $t = 220T_0$. Therefore the ion spectrum from Coulomb explosion is very broad without the wakefield modulation. This result further proves the scheme that the laser wakefield narrows down the ion spectrum and brings the high-quality argon ion beam.

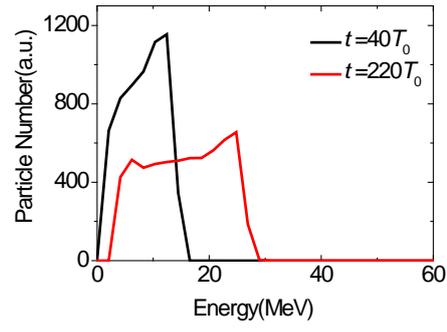

Supplementary Figure-2 Ion spectrum from the laser-induced Coulomb explosion without wakefield modulation.